\documentclass[10pt,twocolumn,letterpaper]{article}

\usepackage[pagenumbers]{cvpr} 

\usepackage{graphicx}
\usepackage{amsmath}
\usepackage{amssymb}
\usepackage{booktabs}
\usepackage[misc]{ifsym}

\usepackage[pagebackref,breaklinks,colorlinks]{hyperref}

\usepackage[capitalize]{cleveref}
\crefname{section}{Sec.}{Secs.}
\Crefname{section}{Section}{Sections}
\Crefname{table}{Table}{Tables}
\crefname{table}{Tab.}{Tabs.}

\def\cvprPaperID{*****} 
\def\confName{CVPR}
\def\confYear{2022}

\usepackage{overpic}
\usepackage{enumitem} 
\usepackage{overpic} 
\usepackage{color}

\definecolor{turquoise}{cmyk}{0.65,0,0.1,0.3}
\definecolor{purple}{rgb}{0.65,0,0.65}
\definecolor{dark_green}{rgb}{0, 0.5, 0}
\definecolor{orange}{rgb}{0.8, 0.6, 0.2}
\definecolor{red}{rgb}{0.8, 0.2, 0.2}
\definecolor{darkred}{rgb}{0.6, 0.1, 0.05}
\definecolor{blueish}{rgb}{0.0, 0.3, .6}
\definecolor{light_gray}{rgb}{0.7, 0.7, .7}
\definecolor{pink}{rgb}{1, 0, 1}
\definecolor{greyblue}{rgb}{0.25, 0.25, 1}

\usepackage{blindtext}

\renewcommand{\paragraph}[1]{\vspace{1em}\noindent\textbf{#1}.}
\begin{document}

\title{LDP-Net: An Unsupervised Pansharpening Network Based on Learnable Degradation Processes}

\author{Jiahui Ni$^1$, Zhimin Shao$^1$,  Zhongzhou Zhang$^1$,  Mingzheng Hou$^1$,
Jiliu Zhou$^1$,  \\Leyuan Fang$^2$,  Yi Zhang$^1$\thanks{Corresponding author}
\\
$^1$Sichuan University, $^2$Hunan University
\\
{\tt\small \{jiahui\_ni, zz\_zhang\}@stu.scu.edu.cn}, \tt\small shaozm\_3@foxmail.com, \\ \tt\small\{houmingzheng, zhoujl, yzhang\}@scu.edu.cn, \tt\small leyuan\_fang@hnu.edu.cn
}

\maketitle
\section{Abstract}
Pansharpening in remote sensing image aims at acquiring a high-resolution multispectral (HRMS) image directly by fusing a low-resolution multispectral (LRMS) image with a panchromatic (PAN) image. The main concern is how to effectively combine the rich spectral information of LRMS image with the abundant spatial information of PAN image. Recently, many methods based on deep learning have been proposed for the pansharpening task. However, these methods usually has two main drawbacks: 1) requiring HRMS for supervised learning; and 2) simply ignoring the latent relation between the MS and PAN image and fusing them directly. To solve these problems, we propose a novel unsupervised network based on learnable degradation processes, dubbed as LDP-Net.  A reblurring block and a graying block are designed to learn the corresponding degradation processes, respectively. In addition, a novel hybrid loss function is proposed to constrain both spatial and spectral consistency between the pansharpened image and the PAN and LRMS images at different resolutions. Experiments on Worldview2 and Worldview3 images demonstrate that our proposed LDP-Net can fuse PAN and LRMS images effectively without the help of HRMS samples, achieving promising performance in terms of both qualitative visual effects and quantitative metrics.

\section{Introduction}
\label{sec:intro}
Nowadays, numerous remote sensing images are obtained to monitor the conditions of agriculture, forestry, ocean, land, environmental protection and meteorology \cite{shao2019deep}. Usually, most earth observation satellites can provide two kinds of images, namely, panchromatic (PAN) images with a high spatial resolution band and multispectral (MS) images with higher spectral resolution but lower spatial resolution, which are limited to the image signal-to-noise ratio (SNR) and data storage and transmission. Naturally, the technique for PAN and MS image fusion has been proposed and developed. This technology, which is known as pansharpening, integrates the complementary advantages of spatial and spectral information respectively from PAN and MS images to obtain high spatial resolution MS images. Fused images with both high spectral and spatial resolution can achieve better results in subsequent tasks, such as image classification and object detection \cite{zhang2013hyperspectral}.

In early research, many traditional methods were proposed to develop pansharpening algorithms, and most of them can be generally summarized into three categories. (1) Methods based on component substitution (CS) \cite{haydn1982application} attempt to transform MS images and PAN images into a new space in which the structural component of MS images can be substituted by PAN images to achieve spatial information injection. Representative attempts include principal component analysis (PCA) \cite{shah2008efficient}, intensity-hue-saturation (IHS) \cite{rahmani2010adaptive}, and Gram-Schmidt adaptive (GSA) transform \cite{aiazzi2007improving}. (2) Multiresolution-analysis-based methods utilize the high frequencies of PAN images to restore the spatial details in MS images. To extract this high-frequency information in PAN images, various transform algorithms are applied, such as Laplacian pyramid transform \cite{shen2016integrated}, discrete wavelet transform (DWT) \cite{pradhan2006estimation}, and support value transform \cite{zheng2007multisource}. (3) Model-based methods \cite{ballester2006variational} treat pansharpening as an inverse process of the degradation in which the ideal high-resolution multispectral (HRMS) image degenerates to a PAN image and low-resolution multispectral (LRMS) image. One typical example is the band-dependent spatial detail (BDSD) method \cite{garzelli2007optimal}. However, these methods exhibit a certain degree of spectral distortions owing to some prior assumptions, which are hard to be generalized to different situations \cite{thomas2008synthesis}.
        
In the past decade, deep learning approaches, especially convolutional neural networks (CNNs), have achieved excellent performance in various fields, including computer vision and image processing tasks \cite{he2016deep}. Some pioneering methods have applied CNNs to the pansharpening task. Typical examples include PNN \cite{masi2016pansharpening}, PanNet \cite{yang2017pannet}, PSGAN \cite{liu2020psgan}, RED-cGan \cite{shao2019residual} and TFNet \cite{liu2020remote}. These supervised learning methods use an end-to-end network to learn the pansharpening process and achieve desirable performance with high spatial resolution and few spectral distortions. However, two vital problems still exist in most CNN-based methods. The first issue is that most networks are based on supervised learning and the training data are generated following Wald’s protocol \cite{wald1997fusion}. These models perform spatial downsampling and blurring operations on the MS images to obtain the LRMS images and treat the original MS images as ground truth. These operations may not be consistent with the degradation processes in the real situation. The other issue is that these schemes do not effectively utilize the rich spatial information of PAN images \cite{ma2020pan} and ignore the relation between MS images and PAN images.
        
To address these problems, we propose a novel unsupervised network for pansharpening based on a two-stream CNN-based architecture with two learnable degradation processes, dubbed as LDP-Net. Pansharpening can be regarded as a super-resolution or deblurring problem \cite{zhong2016remote} with additional PAN images and aims to restore the spatial details from PAN images and simultaneously maintain the spectral information of LRMS images. Owing to the lack of ground truth, the inverse process of pansharpening can be divided into two degradation processes: one process uses a spectral response function to transform the HRMS image into a single grayed image similar to the PAN image, and the other process models a spatial blurring operation from the HRMS image into an upsampled LRMS image with a blurring kernel. In the proposed LDP-Net, we adopt two CNN modules to learn two degradation processes. Moreover, according to the relation between MS and PAN images, we propose a new loss function to effectively constrain both spatial and spectral information. Furthermore, a KL divergence loss function is proposed to maintain the spectral distribution of the difference between the MS and PAN images at two resolutions, which has never been explored. As a result, our proposed model achieves desirable performance in that the predicted HRMS image can preserve the high spatial resolution of the PAN image and rich spectral information of the LRMS image under unsupervised conditions. The main contributions of this paper are summarized as follows.

\paragraph{Contributions}
\begin{itemize}[leftmargin=*]
\setlength\itemsep{-.3em}
\item  An unsupervised pansharpening model is proposed based on a two-stream end-to-end network, which is trained without relying on supervised labels. The model can easily converge with a specific training scheme.
\item Different from other models with specified degradation operators, our proposed model learns the degradation processes in a data-driven manner.
\item A novel hybrid loss function, which consists of three parts, is proposed. The first two parts maintain the spatial and spectral consistency between the inputs and the predicted HRMS image in two different resolutions. The other part constrains the difference between the MS and PAN images at different resolutions to have similar distributions.
\item Extensive experiments on different remote sensing datasets demonstrate the effectiveness and robustness of our method over several state-of-the-art methods in both qualitative and quantitative aspects.
\end{itemize}

The remainder of this paper is organized as follows. In Section \ref{sec:related}, we review related works on pansharpening. Section \ref{sec:method} introduces the framework of the proposed unsupervised model and the loss function for training without labels. In Section \ref{sec:results}, extensive experiments were conducted to illustrate our pansharpening method compared with several representative traditional, supervised and unsupervised learning based approaches. Finally, conclusion is drawn in Section \ref{sec:conclusions}.
\section{Related works}
\label{sec:related}
Numerous pansharpening methods have emerged in recent decades, and this section briefly reviews these methods, including classic approaches, supervised learning based approaches and unsupervised learning based approaches.

\subsection{\paragraph{Classic Methods}}
Traditional pansharpening methods can be roughly classified into three categories. First, early pansharpening studies focused on CS. Some components of the upsampled LRMS images are substituted by corresponding components of PAN images in a specific transform domain. The spectral information and spatial information are separated using a simple and fast transformation, such as IHS \cite{rahmani2010adaptive}, principal components transform \cite{chavez1991comparison} and GSA transform \cite{aiazzi2007improving}. Moreover, Dou \emph{et al.} \cite{dou2007general} proposed a general framework to implement these CS-based methods systematically. These methods can effectively achieve high spatial resolution but may cause spectral distortions in the pansharpened results. The second category is multiresolution-analysis-based methods, which apply multiscale decomposition techniques to inject high-frequency information of the PAN image into the upsampled LRMS image. High-frequency spatial information is usually extracted by several transform algorithms, such as wavelet transform \cite{otazu2005introduction}, Laplacian pyramid transform \cite{shen2016integrated}, curvelet transform \cite{nencini2007remote} and contourlet transform \cite{shah2008efficient}. Although these methods can achieve improved performance in spectral fidelity, they may also cause aliasing distortion and blurring effects in spatial details. The third type is model-based methods. For instance, Garzelli \emph{et al.} \cite{garzelli2007optimal} presented two linear injection models, including the single spatial detail (SSD) model and the BDSD model and optimized the models by minimizing the squared error between the original MS image and pansharpened results. Another pansharpening model proposed by Wright achieved fast image fusion with a Markov random field \cite{wright1999fast}. In addition, \cite{guo2014online} adopted an online coupled dictionary learning approach to model the relation between LRMS and PAN images to reduce the spectral distortion and restore the spatial details.

\subsection{\paragraph{Supervised Learning Based Approaches}}
These deep learning methods specifically design a CNN-based network driven by large quantities of paired training data and achieve better performance than traditional methods. Motivated by the super-resolution convolutional neural network (SRCNN) model \cite{dong2015image}, Giuseppe \emph{et al.} \cite{masi2016pansharpening} first proposed a three-layer CNN-based network named PNN according to the characteristics of remote sensing images. Later, Yang \emph{et al.} \cite{yang2017pannet} directly added the upsampled LRMS image to the output of the network to maintain spectral consistency and treated the edges of PAN and LRMS images as the inputs of the network to restore the spatial details. However, introducing only high-frequency information and superimposing the upsampled LRMS image on the results can cause a blurring effect and lead the training difficult to converge. Scarpa \emph{et al.} \cite{scarpa2018target} adopted a target-adaptive usage modality to ensure that a lightweight network can be applied to different remote sensing sensors. In the deep residual pansharpening neural network (DRPNN) model \cite{wei2017boosting}, the concept of residual learning is introduced to form a very deep convolutional neural network, which can further improve the pansharpening performance. Recently, Liu \emph{et al.} \cite{liu2020remote} also incorporated residual learning into a two-stream CNN architecture to fuse the features extracted from both MS and PAN images. Moreover, several generative adversarial network (GAN)-based methods have been proposed to utilize a discriminator to distinguish the generated images from the ground-truth images. In PSGAN \cite{liu2020psgan}, the authors first attempted to produce high-quality pansharpened images with GANs and design a two-stream fusion architecture as the generator and a fully convolutional network as the discriminator. In RED-cGAN \cite{shao2019residual}, a residual encoder-decoder conditional GAN was proposed to produce more details with sharpened images. However,  as we mentioned above, these methods require HRMS images for supervised learning and still suffer from spectral distortions or blurring effects.

\subsection{\paragraph{Unsupervised Learning Based Approaches}}
To address the unreality of simulated data and bridge the gap between classic and supervised learning based approaches, some unsupervised learning based approaches have been developed. Ma \emph{et al.} \cite{ma2020pan} achieved unsupervised pansharpening using one generator and two discriminators that were designed to distinguish the spatial and spectral characteristics between generated and real images, respectively. Motivated by some priors about downsampling and blurring, several methods have been developed for unsupervised pansharpening. For instance, a deep learning prior based on spatial downsampling with blurring has been applied for image fusion to obtain the loss function in \cite{uezato2020guided}. The authors embedded the semantic features extracted from the guidance PAN image by an encoder-decoder network into another deep decoder to generate an output image. Similarly, Luo \emph{et al.} \cite{luo2020pansharpening} designed an iterative network architecture with a PAN-guided strategy and a set of skip connections to continuously extract and fuse the features from the input and then used a fixed unidimensional Gaussian kernel to obtain a blurred version from the fused HRMS image. However, these prior-based methods are limited to handcrafted training data and cannot be effectively applied to real scenes.

In this paper, we propose an unsupervised learning model based on a two-stream CNN network incorporated with two learnable degradation modules that can be adaptive to complex simulated and real situations. Moreover, we specifically design a hybrid spectral loss to effectively maintain spectral consistency between the output and input LRMS images.
\begin{figure*}[!t]
\begin{center}
\includegraphics[width=\linewidth]{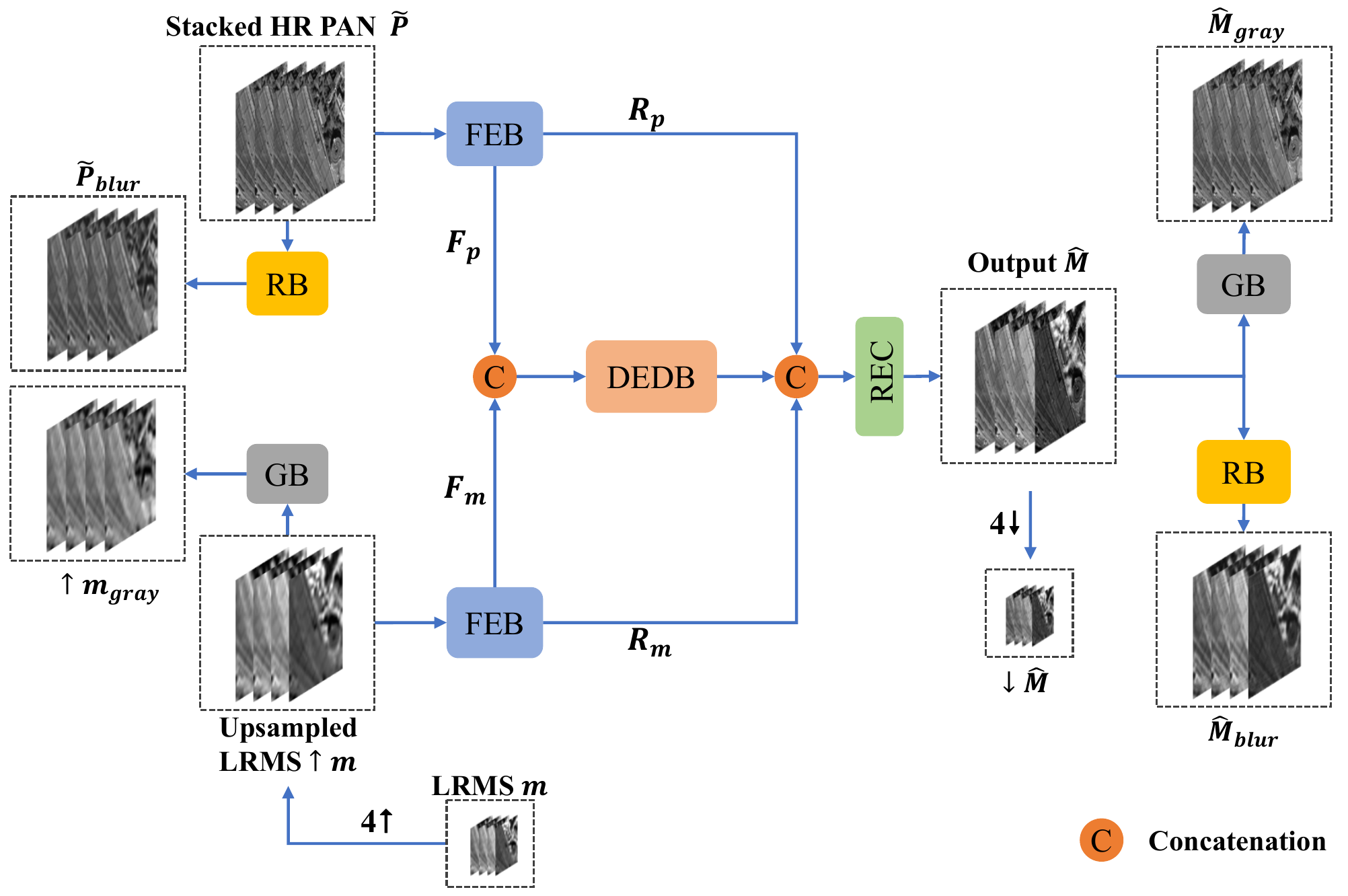}
\end{center}
\vspace{-0.4cm}
\caption{Overview of the proposed LDP-Net for pansharpening. FEB denotes the feature extraction block. DEDB denotes the dense encoder-decoder block. RB and GB represent the reblurring block and graying block, respectively. REC stands for the reconstruction block. $4 \uparrow$ and $4 \downarrow $ stand for 4 times upsamling and downsampling, respectively.  $F$ and $R$  denote the shallow features and residual connection, respectively.}
\label{figure1}
\end{figure*} 
\section{Method}
\label{sec:method}
\subsection{\paragraph{Problem Formulation and Framework}}
\label{sec:method1}
Unsupervised pansharpening aims to obtain the pansharpened HRMS image by fusing the LRMS image and the HR PAN image without ground-truth label. We denote the LRMS image by $m \in R^{w\times h\times C}$, the corresponding HR PAN image by $P \in R^{W\times H}$, the pansharpened HRMS image by $\widehat M \in {R^{W \times H \times C}}$ and the ground-truth HRMS image by $ M \in {R^{W \times H \times C}}$. $W$ and $H$ represent the width and height of high-resolution images, respectively, while $w$ and $h$ represent the width and height of low-resolution images, respectively. $C$ is the number of spectral bands of the multispectral image and usually, $C=4$. The scale factor for spatial resolution ratio is defined  as $r=W/w=H/h$ and usually $r=4$.

Our proposed LDP-Net is based on a two-stream encoder-decoder fusion network. As shown in Fig. \ref{figure1}, the network mainly consists of several different modules, including feature extraction block (FEB), dense encoder-decoder block (DEDB), reconstruction block (REC), graying block (GB) and reblurring block (RB). First, we interpolate the LRMS image $m$ to the upsampled LRMS image $\uparrow m \in {R^{W \times H \times C}}$ with same resolution as that of the PAN image and copy the single-band PAN image $C$ times to form a $C$-band tensor $\widetilde P \in {R^{W \times H \times C}}$. Then, $\uparrow m$ and $\widetilde P$ are fed into FEB to obtain the shallow spectral and spatial features $F_m$ and $F_p$, respectively. Then, we use DEDB \cite{huang2017densely}, which has a strong inference ability to further extract and fuse the deep features. Finally, the predicted HRMS image $\widehat M$, is reconstructed from the concatenation of the deep features and shallow features via two residual connections \cite{he2016deep}. The fusion process takes the following general form:
\begin{equation}
\widehat M = f(\uparrow m,\widetilde P;\Theta),
\label{1}
\end{equation}
where $f\left(  \cdot  \right)$ is the two-stream encoder-decoder fusion model, which takes $\uparrow m$ and $\widetilde P$ as the inputs and generates the desired HRMS image $\widehat M$, while $\Theta$ is the collection of parameters for this model.

Since we do not have the HRMS image as labels, to achieve unsupervised learning, two degradation processes, namely, the degradation between the ideal HRMS image $M$ and the HR PAN image $P$ and the degradation between the ideal HRMS image $M$ and the upsampled LRMS image $\uparrow m$, are formulated to add extra constraints on the training procedure of $f$ as follows \cite{garzelli2016review}:
\begin{equation}
P = \sum\limits_{i = 1}^C {{\alpha _i}{M_i}},
\label{2}
\end{equation}
and
\begin{equation}
\uparrow m = k * M,
\label{3}
\end{equation}
where $M_i$ denotes the $i$-th band of the ideal HRMS image, $\alpha_i$ is the corresponding weighting coefficient and $k$ represents the spatial blur kernel. (\ref{2}) can be regarded as the degradation process of graying an MS image, which is similar to graying a RGB image, while (\ref{3}) can be regarded as the blurring process. Inspired by the forms of (\ref{2}) and (\ref{3}), a channel attention module \cite{hu2018squeeze} is adopted to simulate the graying degradation as GB and a convolution module is employed to simulate the blurring degradation as RB. The parameters of both modules can be learned from training data. Consequently, our model can be optimized by minimizing the loss between the inputs and two degraded versions of the output HRMS image $\widehat M$. In addition, we apply the two degradation blocks for the two inputs to obtain their corresponding degraded versions at low resolution. It is worth mentioning that the different GBs and RBs used in our proposed LDP-Net share the same parameters, respectively.

\begin{figure*}[t]
\begin{center}
\includegraphics[width=\linewidth]{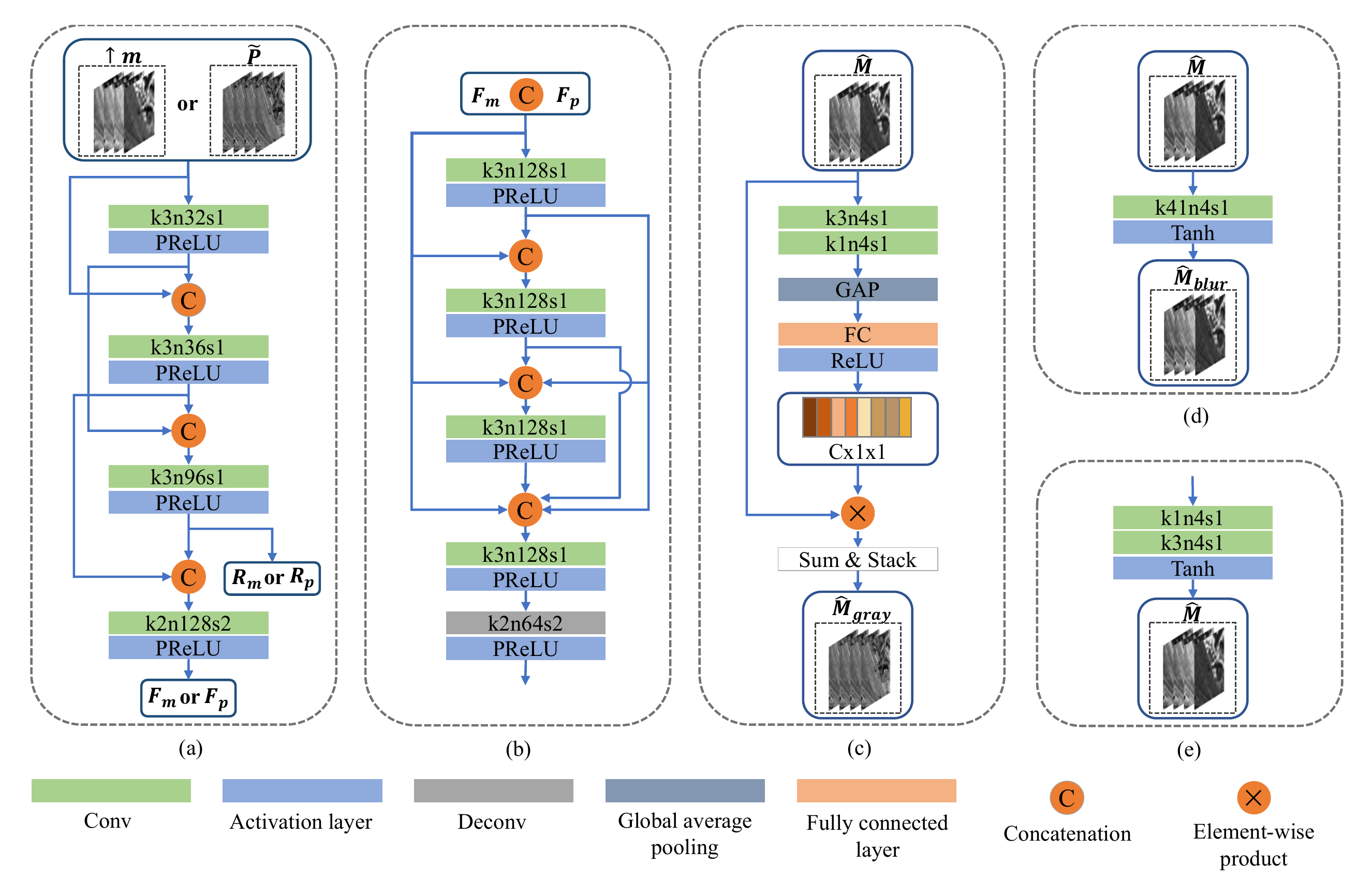}
\end{center}
\vspace{-0.4cm}
\caption{The structure of (a) FEB, (b) DEDB, (c) GB,  (d) RB, and (e) REC, where k3n128s1 denotes a convolution layer with a 3×3 kernel size, 128 channels and stride 1.}
\label{figure2}
\end{figure*} 

\subsection{\paragraph{Loss Function}} 
Given the upsampled LRMS image $\uparrow m$ and the stacked HR PAN image $\tilde P$ as the inputs, our network produces the desired HRMS image $\widehat M$ and four degraded images using the learned degradation operations, which are respectively defined as follows:
\begin{equation}
{\widehat M_{gray}} = G\left( {\widehat M} \right),
\end{equation}
\begin{equation}
{\widehat M_{blur}} = B\left( {\widehat M} \right),
\end{equation}
\begin{equation}
\uparrow {m_{gray}} = G\left( { \uparrow m} \right),
\end{equation}
\begin{equation}
{\widetilde P_{blur}} = B\left( {\widetilde P} \right),
\end{equation}
where ${\widehat M_{gray}} \in {R^{W \times H \times C}}$ is the grayed version of $\widehat M$, ${\widehat M_{blur}} \in {R^{W \times H \times C}}$ is the blurred version of $\widehat M$, $ \uparrow {m_{gray}} \in {R^{W \times H \times C}}$ denotes the grayed version of $\uparrow m$ and ${\widetilde P_{blur}} \in {R^{W \times H \times C}}$ denotes the blurred version of $\widetilde P$. $G\left(  \cdot  \right)$ and $B\left(  \cdot  \right)$ represent the corresponding degradation functions of GB and RB, respectively. Then, our model utilizes these degraded versions to calculate the loss without the ground truth. The proposed loss function contains three parts: spatial loss, spectral loss and spectral KL divergence loss.

\emph{1}) Spatial Loss : The degradation relationship between the MS image and PAN image can be used to restore the high-resolution spatial information of the output HRMS image. Thus, the spatial loss of our method, which can be divided into spatial constraints at both low and high resolutions, is defined as:
\begin{equation}
{L_{spatial}} = \left\| {{{\widetilde P}_{blur}} -  \uparrow {m_{gray}}} \right\|_2^2+ \delta  * \left\| {\widetilde P - {{\widehat M}_{gray}}} \right\|_2^2 ,
\label{4}
\end{equation}
where ${\left\|  \cdot  \right\|_2}$ denotes the L2 norm and $\delta$ represents a regularization parameter to balance the two terms. The first term represents the spatial constraint at low-resolution , and the second term represents the spatial constraint at high-resolution after upsampling. The proposed spatial loss devotes to ensuring the consistency of spatial information extracted by two degradation modules at different resolutions.

\emph{2}) Spectral Loss : Another degradation between the HRMS image and the upsampled LRMS image can be regarded as the blurring operation, which can be used to maintain the spectral consistency between the output HRMS image and the input upsampled LRMS image at different resolutions. Then, similar to (\ref{4}), the spectral loss is defined as:
\begin{equation}
{L_{spectral}} = \left\| { \uparrow m - {{\widehat M}_{blur}}} \right\|_{_2}^2{\rm{ + }}\gamma  * \left\| {m -  \downarrow \widehat M} \right\|_2^2,
\end{equation}
where $\gamma$ denotes a regularization parameter to balance the two terms.

\emph{3}) Spectral KL Divergence Loss : On the other hand, we consider the inverse process of graying degradation and note that the spectral information of MS images in different spectral bands should follow a specific pattern. The difference between the MS image and PAN image at different resolutions should have similar distributions. Based on this consideration, we use the softmax function to transform the residual terms into a probability distribution form. Then, the spectral Kullback–Leibler (KL) divergence loss is added to regularize the distribution of the residual terms at different resolutions, which is formulated as follows:
\begin{equation}
{L_{KL}} = KL(p({x_{low}})\left\| {q(x))} \right.,
\end{equation}
where $p({x_{low}}) = softmax ( \uparrow m -  \uparrow {m_{gray}})$ and $q(x) = softmax (\widehat M - \widetilde P)$. ${x_{low}} =  \uparrow m -  \uparrow {m_{gray}}$ denotes the residual features between the MS image and the PAN image at low resolution and $x = \widehat M - \widetilde P$ stands for the residual features between the MS image and the PAN image at high resolution. We reshape the residual terms into a one-dimensional vector and apply the softmax function to rescale the elements. Then KL divergence is applied to impose both terms have similar distributions. The spectral KL divergence loss can effectively reduce the spectral artifacts in the fused results, which will be demonstarted in the experimental section.

In summary, we utilize spatial loss and spectral loss to simultaneously restore the spatial details and preserve the spectral information from the inputs. Moreover, an additional spectral KL divergence loss is proposed to further adjust the spectral qualities. Finally, our proposed unsupervised model is trained by minimizing the following loss function:
\begin{equation}
L = \alpha {L_{spatial}} + \beta {L_{spetral}} + \mu {L_{KL}}.
\end{equation}
where $\alpha$, $\beta$ and $\mu$ are the weights that are empirically set in our experiments. It can be seen that the proposed loss function can be used to train the proposed LDP-Net without the HRMS image (ground truth) via two degradation processes that can learn the latent characteristics of the output HRMS image.

\subsection{\paragraph{Network Architecture}}
As mentioned in Subsection \ref{sec:method1}, there are several CNN-based blocks that are designed to implement our proposed network framework, including FEB, DEDB, GB and RB. Specifically, FEB is used to extract the shallow features from the upsampled LRMS image and HR PAN image to contribute to the subsequent fusion step. Thus, given $\uparrow m$ or $\widetilde P$ as the inputs, the corresponding shallow features $F_m$ or $F_p$ can be obtained as:
\begin{equation}
F_m = {f_{FEB}}( \uparrow m),
\end{equation}
and
\begin{equation}
F_p = {f_{FEB}}(\widetilde P),
\end{equation}
where $ {f_{FEB}}$ represents the operation of FEB. It must mention that both blocks have the same structure but different parameters that extract different features from the MS and PAN images, respectively. As shown in Fig. \ref{figure2}(a), three convolutional layers with several adjacent residual connections are adopted to extract features from different depths and one downsampling convolutional layer is used to reduce the size of features. As shown in Fig. \ref{figure1}, $R_p$ and $R_p$ denote the output of the residual connection from the PAN image and upsampled LRMS image, respectively. All convolutional layers are followed by PReLU activation function. The extracted features $F_m$ and $F_p$ are concatenated as the input of the subsequent DEDB.

The role of DEDB is to learn more high-level features and fuse sufficient spatial and spectral information. As shown in Fig. \ref{figure2}(b), we adopt four convolutional layers with dense connections to enhance the fusion and inference abilities. Then, the fused features are fed into a deconvolutional layer for upsampling before concatenation with the two residual connections. To reconstruct the output HRMS image, we use a reconstruction block (REC) that consists of two convolutional layers followed by a \emph{tanh} activation layer as demonstrated in Fig. \ref{figure2}(e).

GB and RB are vital parts of our proposed unsupervised model. Taking the output HRMS image or the upsampled LRMS image as the input, GB is implemented aided by the channel attention mechanism, as shown in Fig. \ref{figure2}(c). First, we adopt two convolutional layers to transform the input into weight features and use global average pooling (GAP) and fully connected layers to obtain the channel weight vector, which is used to simulate the graying process. Finally, we obtain the stacked output by copying it in the channel dimension. For RB, we implement this operation using a single convolution layer as illustrated in Fig. \ref{figure2}(d).

\section{Results}
\label{sec:results}
\subsection{Experimental Setup}
\label{sec:result1}
\emph{1}) Datasets and Metrics: To evaluate the performance of the proposed method, we conduct experiments on two datasets: Worldview2 (WV2) and Worldview3 (WV3). The spatial resolutions of the MS and PAN images for WV2 satellite are 1.84 m and 0.46 m, respectively, while those for WV3 satellite are 1.24 m and 0.31 m. Both satellites have four bands: red, green, blue and near-infrared. Moreover, our proposed model was trained on the preprocessed images after linear stretching rather than the original source data. Then, we produced the training data following Wald’s protocol \cite{wald1997fusion}, cropped the PAN and LRMS images into 7224 and 5820 patch pairs of sizes 128 × 128 and 32 × 32 without overlap in WV2 and WV3, respectively, and randomly split them into 90$\%$ and 10$\%$ as our training data and validation data, respectively. Furthermore, another 324 and 164 pairs were selected to implement test experiments of the reduced resolution and full resolution from WV2 and WV3, respectively.

The performance of different methods in the reduced-resolution and full-resolution experiments are evaluated by different quantitative metrics. In reduced-resolution testing, four widely used metrics with reference are involved, namely, the spectral angle mapper (SAM) \cite{yuhas1992discrimination}, spatial correlation coefficient (SCC) \cite{zhou1998wavelet}, relative global synthesis errors (ERGAS) \cite{wald2000quality} and 4-band extension of the universal image quality index (Q4) \cite{alparone2004global}, while the quality with no-reference (QNR) \cite{alparone2008multispectral} and its spectral components ${D_\lambda }$ and spatial components $D_S$ are used in full-resolution testing.

\emph{2}) Implementation Details: No postprocessing operations were applied on the output HRMS image. The network was trained with approximately 50 epochs. The Adam optimizer \cite{kingma2014adam} was used to minimize the loss function, with an initial learning rate of 1e-4, and it was decayed by 0.1 every 10 epochs. The batch size was set to 16, the weight of loss $\alpha$ was set to 1, $\beta$ was set to 1, $\mu$ was set to 0.01, $\delta$ was set to 10 and $\gamma$ was set to 20. The network was implemented in PyTorch and trained on an Nvidia GeForce GTX 1080Ti GPU. The codes for this work can be downloaded at https://github.com/suifenglian/LDP-Net.

\emph{3}) Comparison Methods: In our experiments, we compared the proposed LDP-Net with several state-of-the-art methods, including PCA \cite{shah2008efficient}, IHS \cite{rahmani2010adaptive}, Brovey \cite{gillespie1987color}, GS \cite{laben2000process}, BSBD \cite{garzelli2007optimal}, additive wavelet luminance proportional (AWLP) \cite{kim2010improved}, PanNet \cite{yang2017pannet}, PNN \cite{masi2016pansharpening} and Pan-GAN \cite{ma2020pan}. The first six methods belong to traditional methods. PNN and PanNet are supervised learning based methods. Pan-GAN is a recently proposed unsupervied method. These methods were reimplemented with the PyTorch framework according to their publicly available codes and retrain all these methods using the same training dataset in our experiments.

\subsection{Comparison at Reduced Resolution}

The experiment was performed on two datasets at reduced resolution, which follows Wald’s protocol. As a result, the original MS image can be used as the reference for evaluations. Figs. \ref{figure3}-\ref{figure6} show four examples cropped from the results of WV2 and WV3. In each case, one region that is marked by a red rectangle is magnified to visualize the differences of these results. In Figs. \ref{figure3}-\ref{figure6}, it can be observed that the results of traditional methods exhibit obvious blurred details and some spatial artifacts. Especially for Brovey, the results suffer from severe spectral distortions and some blurring effects in Figs. \ref{figure3} and \ref{figure4}. For supervised learning based methods, the results of PanNet show more blurring effects and spectral distortions, while PNN can restore the results without noticeable blurring effects and spectral distortions in Figs. \ref{figure3}(i) and \ref{figure4}(i), but spectral distortions can be noticed in Figs. \ref{figure3}(i) and \ref{figure4}(i). For the unsupervised method, Pan-GAN contains rich spatial details but some spatial artifacts and obvious spectral distortions can be noticed. Generally, GAN-based methods are difficult to train and easily generate spatial and spectral artifacts. Compared to other methods, it can be seen that our LDP-Net effectively recovers spatial details and preserves spectral information without introducing any artifacts and the fusion results are more vivid and much closer to the ground truth than other methods, as shown in the magnified regions in Figs. \ref{figure3}(k)-\ref{figure6}(k).

Tables \ref{table1} and \ref{table2} show the average values of the quantitative results of different methods on both WV2 and WV3 datasets. The best results are highlighted in bold. Compared with other traditional and unsupervised methods, the proposed method achieved the best scores in most metrics. Among the CNN-based methods, the proposed method can approach the performance of supervised methods. In particular, our model achieves the best SCC, ERGAS and Q4 scores for WV3 dataset in all the methods, which verifies that our proposed method can effectively fuse the spatial and spectral information without the reference.
\begin{table}[htbp]
\centering
\resizebox{\linewidth}{!}{ 
\begin{tabular}{ccccc}
\hline
            & SAM     & SCC    & ERGAS  & Q4     \\
Ideal value & 0       & 1      & 0      & 1      \\
\hline
PCA         & 12.2933 & 0.8034 & 4.4811 & 0.8956 \\
IHS         & 10.4043 & 0.8566 & 4.2864 & 0.8938 \\
Brovey      & \textbf{10.2148} & 0.8451 & 4.4518 & 0.8699 \\
GS          & 10.4551 & 0.8528 & 4.2579 & 0.8802 \\
BDSD        & 13.5506 & 0.8560 & 3.7855 & 0.8522 \\
AWLP        & 12.0686 & 0.8280 & 3.5772 & 0.8635 \\
Pan-GAN     & 13.5112 & 0.8660 & 3.9333 & 0.9643 \\
Ours        & 12.1264 & \textbf{0.8730} & \textbf{3.4834} & \textbf{0.9853} \\
\hline
PNN         & 10.9829 & 0.9067 & 3.3169 & 0.9432 \\
PanNET      & 11.2514 & 0.8527 & 4.4121 & 0.9758\\
\hline
\end{tabular}
} 
\caption{Quantitative Results On The Worldview2 Dataset At Reduced Resolution.}
\label{table1}
\end{table}
\begin{table}[htbp]
\centering
\resizebox{\linewidth}{!}{ 
\begin{tabular}{ccccc}
\hline
            & SAM     & SCC    & ERGAS  & Q4     \\
Ideal value & 0       & 1      & 0      & 1      \\
\hline
PCA         & 9.3228  & 0.8817 & 4.2813 & 0.9361 \\
IHS         & 9.0414  & 0.8855 & 4.3394 & 0.9469 \\
Brovey      & \textbf{8.9115}  & 0.8476 & 5.2983 & 0.8509 \\
GS          & 9.2403  & 0.8824 & 4.3472 & 0.9251 \\
BDSD        & 13.2167 & 0.8942 & 3.4667 & 0.8942 \\
AWLP        & 9.4067  & 0.8760 & 3.1305 & 0.9230 \\
Pan-GAN     & 12.7301 & 0.9199 & 3.2724 & 0.9689 \\
Ours        & 10.2786 & \textbf{0.9417} & \textbf{2.4232} & \textbf{0.9820} \\
\hline
PNN         & 10.0536 & 0.9411 & 2.6066 & 0.9732 \\
PanNET      & 10.3519 & 0.8715 & 4.6296 & 0.9761 \\
\hline
\end{tabular}
} 
\caption{Quantitative Results On The Worldview3 Dataset At Reduced Resolution.}
\label{table2}
\end{table}

\subsection{Comparison at Full Resolution}

In this section, all the methods were validated on real data. Figs. \ref{figure7}-\ref{figure10} illustrate the representative pansharpened results for the real WV2 and WV3 data. Moreover, to verify the robustness of the proposed LDP-Net, the models trained with reduced images were used for the full-resolution test, which means we do not need to train new models for the full-resolution dataset. In these cases, most traditional methods can significantly restore the spatial information compared with that in LRMS images but still suffer from a certain degree of spectral shift. In contrast, AWLP reduces the spectral distortion in the results while still retaining noticeable spatial blurring effects. Compared with these traditional methods, CNN-based models can effectively maintain spectral consistency and improve the spatial resolution over different datasets. However, the results of PanNet still generate sensible blurring effects and artifacts. PNN can achieve better performance and restore the spatial details well with a high spectral resolution, but spectral distortions are still observed in partial regions, as shown in Fig. \ref{figure7}(j) and \ref{figure9}(j), where the red buildings are not as vividly colored as those obtained by other methods. Pan-GAN, which achieves unsupervised learning using spatial and spectral discriminators, can improve the spatial and spectral resolution but still introduces some artifacts and spectral distortions to the results. However, it is obvious that in the magnified regions indicated by red boxes, our proposed method preserves better spatial details and maintains higher spectral consistency than other methods. Apparently, our pansharpened images are clearer and more vivid than all the other methods, as shown in Figs. \ref{figure7}(l)-\ref{figure10}(l).

Due to lack of ground truth, QNR, ${D_\lambda }$ and $D_S$ are employed as the quantitative metrics to evaluate the performance of the pansharpened results at full resolution. The quantitative results are shown in Table III. As shown in Table \ref{table3}, we notice that the quantitative results are not quite consistent with the results of the visual inspections and the scores of our LDP-Net only are located in the middle of all the methods. This paradox probably lies in that the nonreference assessment metrics are calculated using the LRMS images, PAN image and pansharpened results to assess spectral and spatial distortion. The results with blurring effects tend to achieve better values due to their similarity to the LRMS images, which has also been mentioned in \cite{qu2020unsupervised}. For example, as shown in Figs. \ref{figure7}(i), \ref{figure7}(j), \ref{figure8}(i) and \ref{figure8}(j), the results of PanNet with more obvious blurring effects have better QNR values than PNN. Hence, nonreference metrics are not always suitable to assess the spectral and spatial distortions of pansharpened results, and it is more important to emphasize visual inspection for comparison at full resolution without ground truth.
\begin{table}[htbp]
\centering
\resizebox{\linewidth}{!}{ 
\begin{tabular}{ccccccc}
\hline
            & \multicolumn{3}{c}{WV2}     & \multicolumn{3}{c}{WV3}     \\
            & ${D_\lambda }$       &  ${D_S }$     & QNR     & ${D_\lambda }$  &   ${D_S }$  &   QNR       \\
Ideal value & 0       & 0      & 1      & 0     &  0     &    1     \\
\hline
PCA     & 0.0788 & 0.1379 & 0.7971 & 0.0069 & 0.271  & 0.7539 \\
IHS     & 0.0207 & 0.1171 & 0.865  & \textbf{0.0044} & 0.2438 & 0.7529 \\
Brovey  & 0.0298 & 0.1016 & 0.875  & 0.0852 & 0.2034 & 0.7328 \\
GS      & \textbf{0.0241} & 0.1171 & 0.8621 & 0.0071 & 0.2409 & 0.7536 \\
BDSD    & 0.0422 & \textbf{0.0569} & \textbf{0.9034} & 0.03   & \textbf{0.0663} & \textbf{0.9061} \\
AWLP    & 0.0485 & 0.0768 & 0.8806 & 0.0435 & 0.0969 & 0.8672 \\
Pan-GAN & 0.0487 & 0.1224 & 0.8377 & 0.0374 & 0.247  & 0.7262 \\
Ours    & 0.0352 & 0.1122 & 0.8576 & 0.0297 & 0.1789 & 0.7968 \\
\hline
PNN     & 0.0984 & 0.1747 & 0.7486 & 0.0677 & 0.2733 & 0.6799 \\
PANNET  & 0.0679 & 0.048  & 0.8882 & 0.0303 & 0.0426 & 0.9285 \\
\hline
\end{tabular}
} 
\caption{Quantitative Results At Full Resolution.}
\label{table3}
\end{table}

\subsection{Ablation Study  of Loss function}
\begin{table*}[htbp]
\centering
\resizebox{\linewidth}{!}{ 
\setlength{\tabcolsep}{3mm}{
\begin{tabular}{cccccccccc}
\hline
     \multicolumn{2}{c}{Losses}  & \multicolumn{4}{c}{WV2}     & \multicolumn{4}{c}{WV3}     \\
  ${L_{spatial\_l}}$       &   ${L_{KL}}$         & SAM     & SCC    & ERGAS  & Q4   & SAM     & SCC    & ERGAS  & Q4       \\
\hline
       &      &24.9557 & 0.8283 & 6.0279 & 0.8959 & 20.3333 & 0.9007 & 4.4323 & 0.9407 \\
 $\surd$  &       &26.6451 & 0.7979 & 6.4634 & 0.8955 & 26.3974 & 0.8834 & 5.8735 & 0.9159 \\
       &  $\surd$ &13.2825 & 0.8794 & 3.4166 & 0.954  & 11.1632 & 0.9395 & 2.4523 & 0.9709 \\
 $\surd$  & $\surd$   &\textbf{12.9600}   & \textbf{0.8796} & \textbf{3.3794} & \textbf{0.9793} & \textbf{10.6724} & \textbf{0.9401} & \textbf{2.4012} & \textbf{0.9833} \\
\hline
\end{tabular}}
} 
\caption{Ablation Results With The Loss Functions.}
\label{table4}
\end{table*}
In this subsection, several experiments were conducted to investigate the impacts of each component in our loss function. Based on two learnable degradation processes, the loss functions play an important role in our unsupervised training process. The proposed loss function can be subdivided into four parts, namely, the spatial loss at high resolution  ${L_{spatial\_h}}=\left\| {\widetilde P - {{\widehat M}_{gray}}} \right\|_2^2$, the spatial loss at low resolution ${L_{spatial\_l}}=\left\| {{{\widetilde P}_{blur}} -  \uparrow {m_{gray}}} \right\|_2^2$, the spectral loss ${L_{spectral}}$ and the spectral KL divergence loss ${L_{KL}}$. ${L_{spatial\_h}}$ and  ${L_{spectral}}$ are used as the basic loss components for the unsupervised training, so Table \ref{table4} only shows quantitative results of the ablation study to validate the effectiveness of ${L_{spatial\_l}}$ and the proposed spectral KL divergence loss. In addition, we display the visual results of different combinations of loss components in Fig. \ref{figure11}. It can be seen that the combination of only ${L_{spatial\_h}}$ and ${L_{spectral}}$ cannot achieve satisfactory performance, which still suffers from severe spectral distortions in pansharpened images. Low-resolution spatial loss can reduce the spectral distortions and artifacts, while the spectral KL divergence loss can obviously eliminate spectral artifacts with high spatial resolution. When all of the loss components are included, the pansharpened images have the best quantitative scores and achieve the best spatial and spectral consistency, fully utilizing the rich spatial information of HR PAN images and the relation between MS images and PAN images. These results verify the effectiveness of our proposed hybrid loss function in both qualitative and quantitative aspects.
\begin{table}[htbp]
\centering
\resizebox{\linewidth}{!}{ 
\begin{tabular}{cccc}
\hline
     & \multicolumn{2}{c}{Run time(s)}  & Params(M) \\
     & WV2  & WV3  &    \\
\hline
PCA     & 0.0044  & 0.0046   & {\rm{ -- }} \\
IHS     & 0.00062 & 0.00064  & {\rm{ -- }} \\
Brovey  & 0.00061 & 0.00061 & {\rm{ -- }} \\
GS      & 0.0038  & 0.0037  & {\rm{ -- }} \\
BDSD    & 0.0173  & 0.0179 & {\rm{ -- }}  \\
AWLP    & 0.0224  & 0.0211  & {\rm{ -- }} \\
Pan-GAN & 0.0035  & 0.0057 &  0.092\\
Ours    & 0.0108  & 0.0158  & 3.963\\
PNN     & 0.0034  & 0.0069  & 0.080\\
PANNET  & 0.0044  & 0.0068 &  0.078\\
\hline
\end{tabular}
} 
\caption{Efficiency comparison  with Different Methods When Processing Inputs Of Size 128 $\times$ 128 $\times$ 4.}
\label{table5}
\end{table}
\subsection{Efficiency Study}

In this section, the computational efficiencies of all comparison methods are evaluated. As mentioned in Section \ref{sec:result}, all deep learning based methods were implemented in PyTorch and tested on an Nvidia GeForce GTX 1080Ti GPU, while all traditional methods were implemented in MATLAB R2019b framework on CPU. Table \ref{table5} lists the computational times of different approaches, which are evaluated by averaging the inference time in the testing set at the reduced resolution experiment. Compared with traditional methods, the computational time of our method is at the middle level. The main reason is that our proposed network contains two additional degradation modules and a deeper network structure. Also due to the learned degradation processes, our model is easier to get convergence in the training phase. Generally, in addition to ensuring the superiority of performance, our proposed unsupervised model makes a reasonable tradeoff between model performance and computational cost.

\section{Conclusions}
\label{sec:conclusions}
In this article, we propose an unsupervised pansharpening method based on two learnable degradation processes. The method can adaptively learn the degraded processes with two corresponding CNN-based modules and successfully achieve unsupervised pansharpening. Moreover, we consider the degradation processes at different resolutions and present a novel hybrid loss that can effectively maintain spatial and spectral consistency. Thus, this unsupervised training strategy adequately improves the spatial details and reduces the spectral distortion in the results. Then, extensive experiments were performed on different-resolution images from two datasets, demonstrating the superiority of our proposed method over other state-of-the-art methods. 

{
    \small
    \bibliographystyle{ieee_fullname}
    \bibliography{macros,main}
}

\appendix





\section{Qualitative results}

\begin{figure*}[t]
\begin{center}
\includegraphics[width=6in]{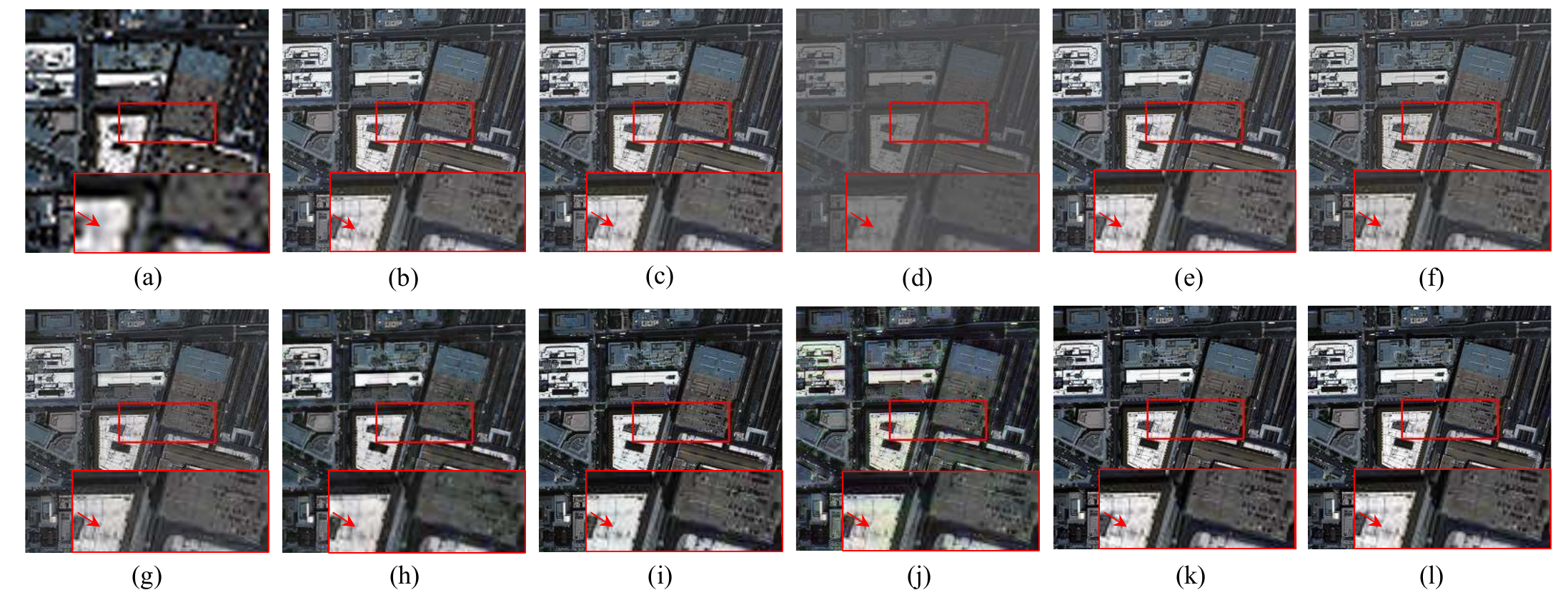}
\end{center}
\vspace{-0.4cm}
\caption{Pansharpened results from different methods on the WV2 dataset at reduced resolution. (a) Upsampled LRMS. (b) PCA. (c) HIS. (d) Brovey. (e) GS. (f) BDSD. (g) AWLP. (h) PanNet. (i) PNN. (j) Pan-GAN. (k) Ours. (l) Ground truth.}
\label{figure3}
\end{figure*}
\vspace{-0.4cm}

\vspace{-0.4cm}
\begin{figure*}[t]
\begin{center}
\includegraphics[width=6in]{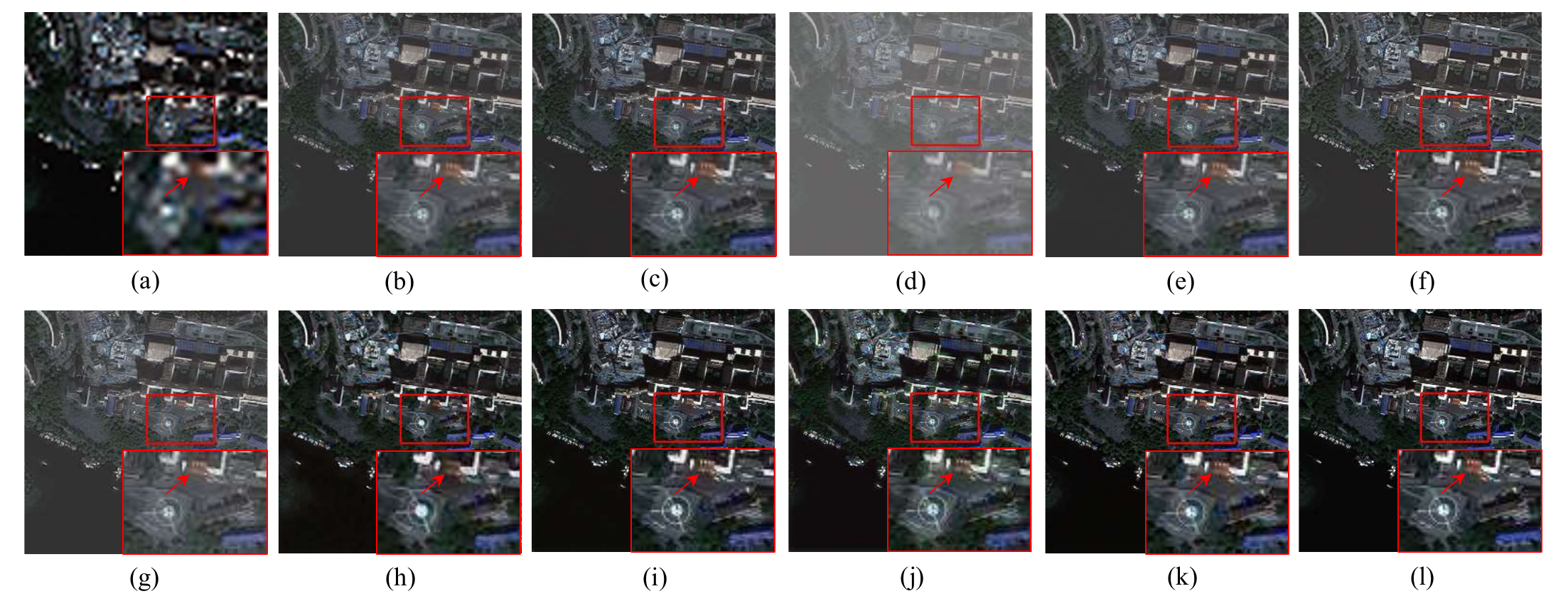}
\end{center}
\vspace{-0.4cm}
\caption{Pansharpened results from different methods on the WV2 dataset at reduced resolution. (a) Upsampled LRMS. (b) PCA. (c) HIS. (d) Brovey. (e) GS. (f) BDSD. (g) AWLP. (h) PanNet. (i) PNN. (j) Pan-GAN. (k) Ours. (l) Ground truth.}
\label{figure4}
\end{figure*}
\vspace{-0.4cm}
\begin{figure*}[t]
\begin{center}
\includegraphics[width=6in]{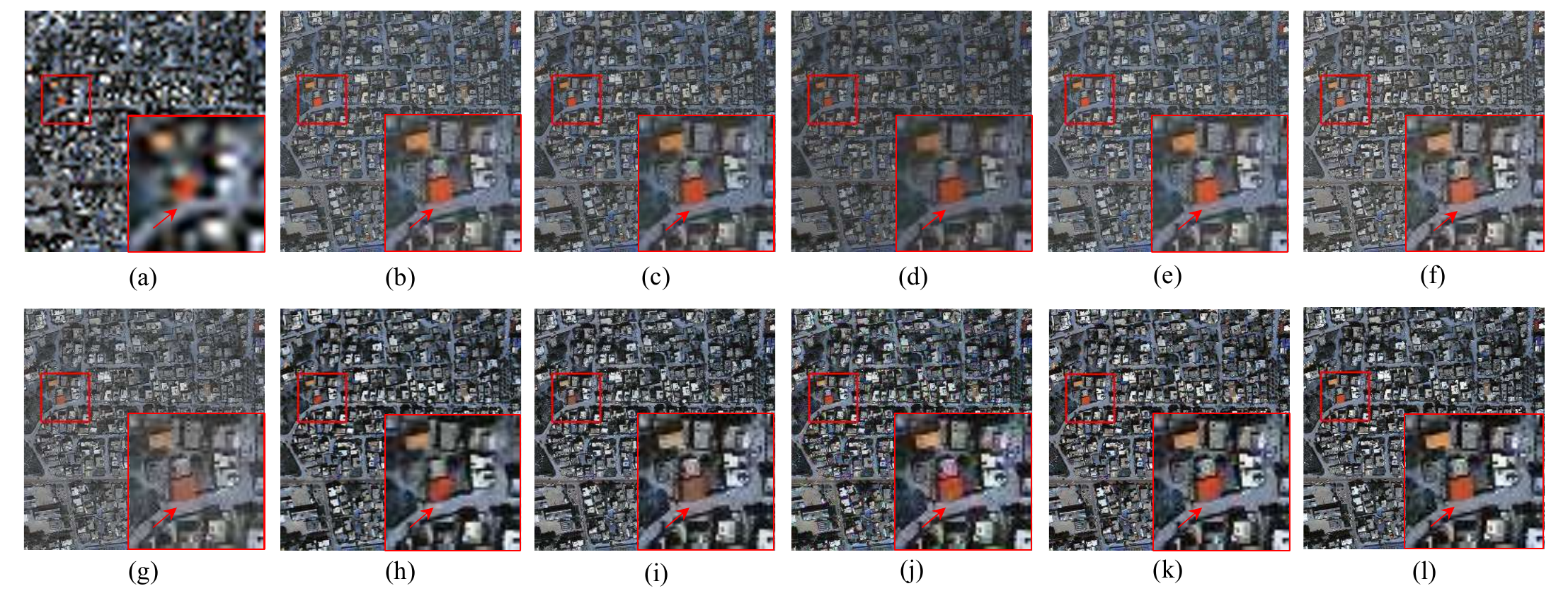}
\end{center}
\vspace{-0.4cm}
\caption{Pansharpened results from different methods on the WV3 dataset at reduced resolution. (a) Upsampled LRMS. (b) PCA. (c) HIS. (d) Brovey. (e) GS. (f) BDSD. (g) AWLP. (h) PanNet. (i) PNN. (j) Pan-GAN. (k) Ours. (l) Ground truth.}
\label{figure5}
\end{figure*}
\vspace{-0.4cm}
\begin{figure*}[t]
\begin{center}
\includegraphics[width=6in]{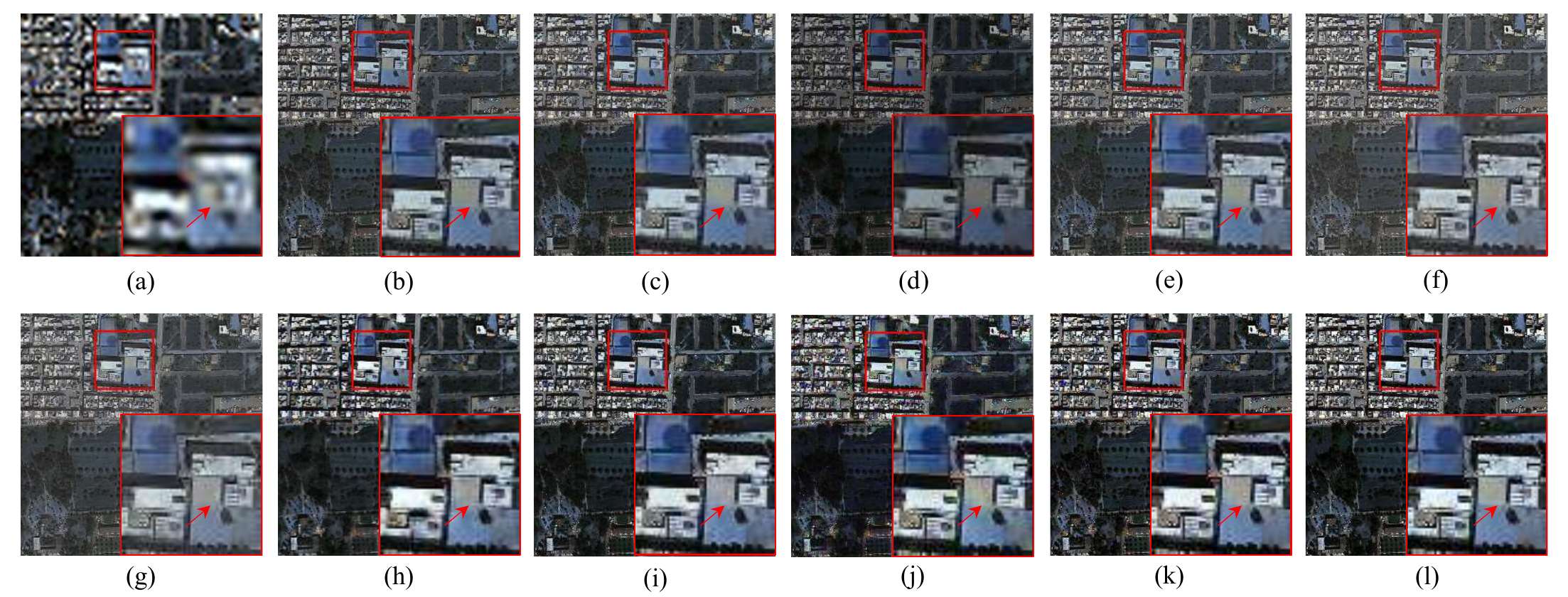}
\end{center}
\vspace{-0.4cm}
\caption{Pansharpened results from different methods on the WV3 dataset at reduced resolution. (a) Upsampled LRMS. (b) PCA. (c) HIS. (d) Brovey. (e) GS. (f) BDSD. (g) AWLP. (h) PanNet. (i) PNN. (j) Pan-GAN. (k) Ours. (l) Ground truth.}
\label{figure6}
\end{figure*}
\vspace{-0.4cm}
\begin{figure*}[t]
\begin{center}
\includegraphics[width=6in]{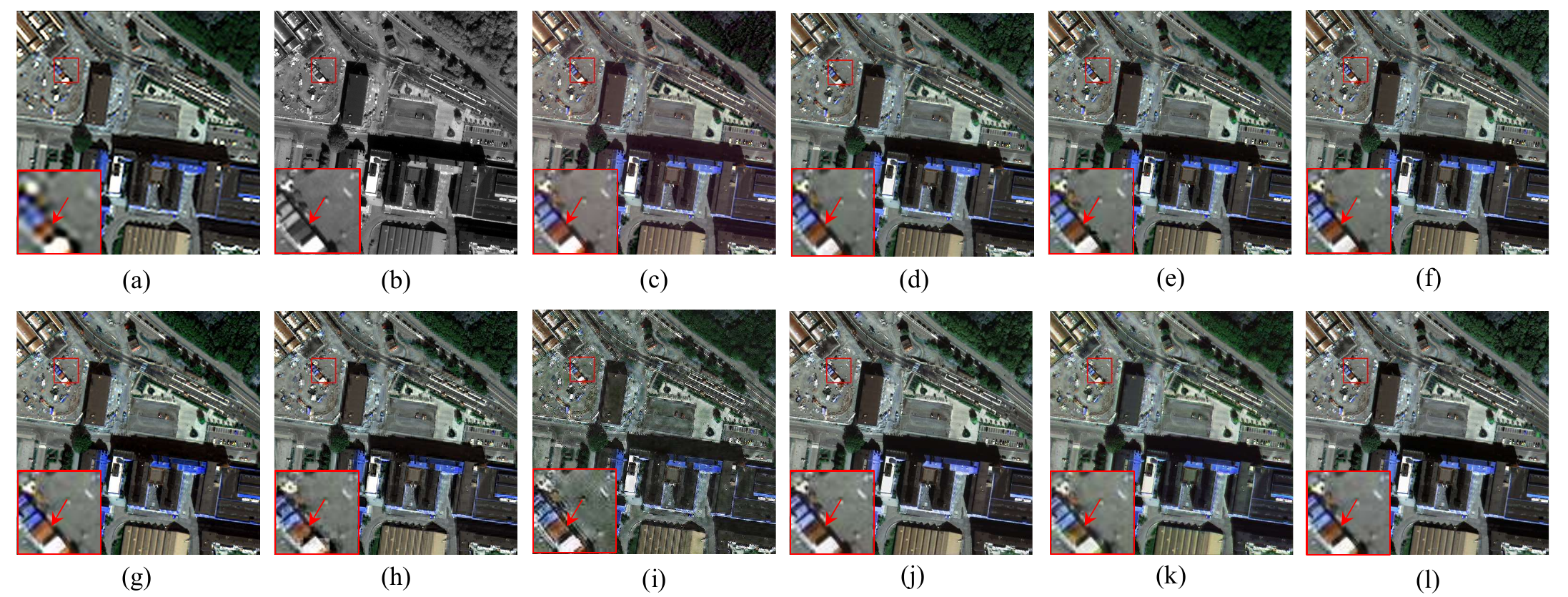}
\end{center}
\vspace{-0.4cm}
\caption{Pansharpened results from different methods on the WV2 dataset at full resolution. (a) Upsampled LRMS. (b) PAN. (c) PCA. (d) HIS. (e) Brovey. (f) GS. (g) BDSD. (h) AWLP. (i) PanNet. (j) PNN. (k) Pan-GAN. (l) Ours.}
\label{figure7}
\end{figure*}
\vspace{-0.4cm}
\begin{figure*}[t]
\begin{center}
\includegraphics[width=6in]{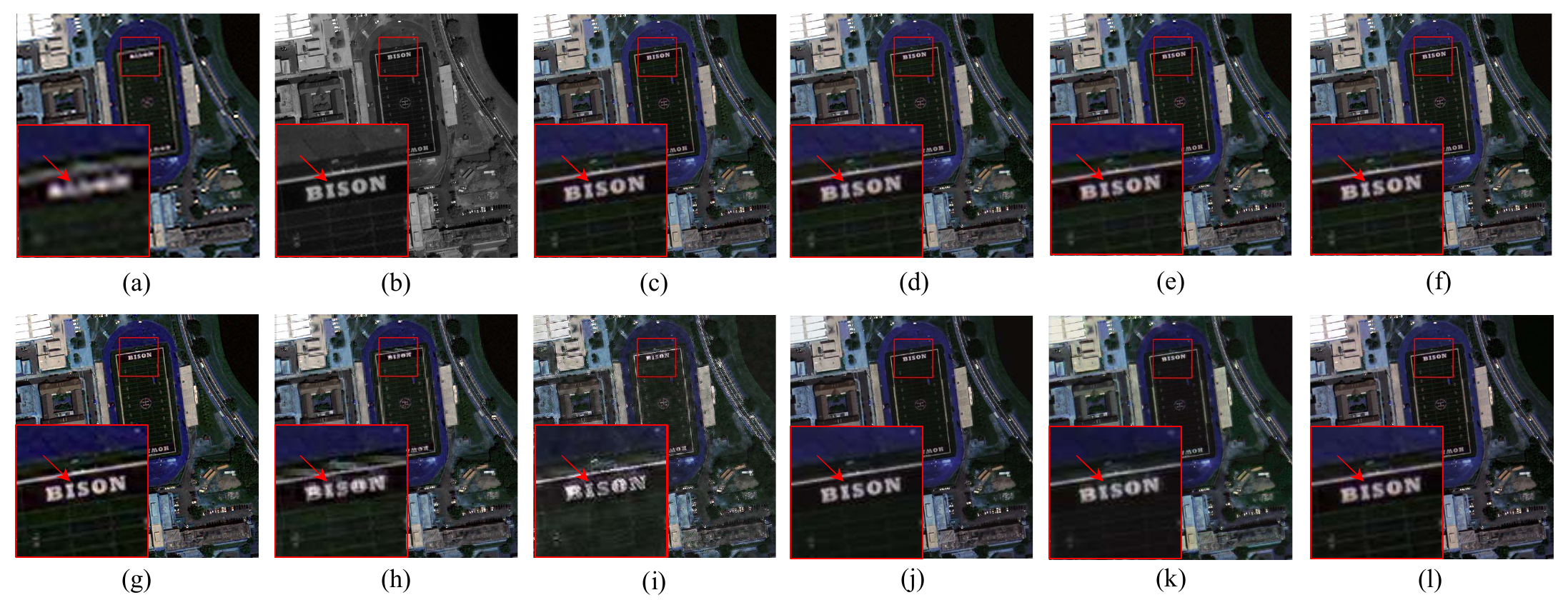}
\end{center}
\vspace{-0.4cm}
\caption{Pansharpened results from different methods on the WV2 dataset at full resolution. (a) Upsampled LRMS. (b) PAN. (c) PCA. (d) HIS. (e) Brovey. (f) GS. (g) BDSD. (h) AWLP. (i) PanNet. (j) PNN. (k) Pan-GAN. (l) Ours}
\label{figure8}
\end{figure*}
\vspace{-0.4cm}
\begin{figure*}[t]
\begin{center}
\includegraphics[width=6in]{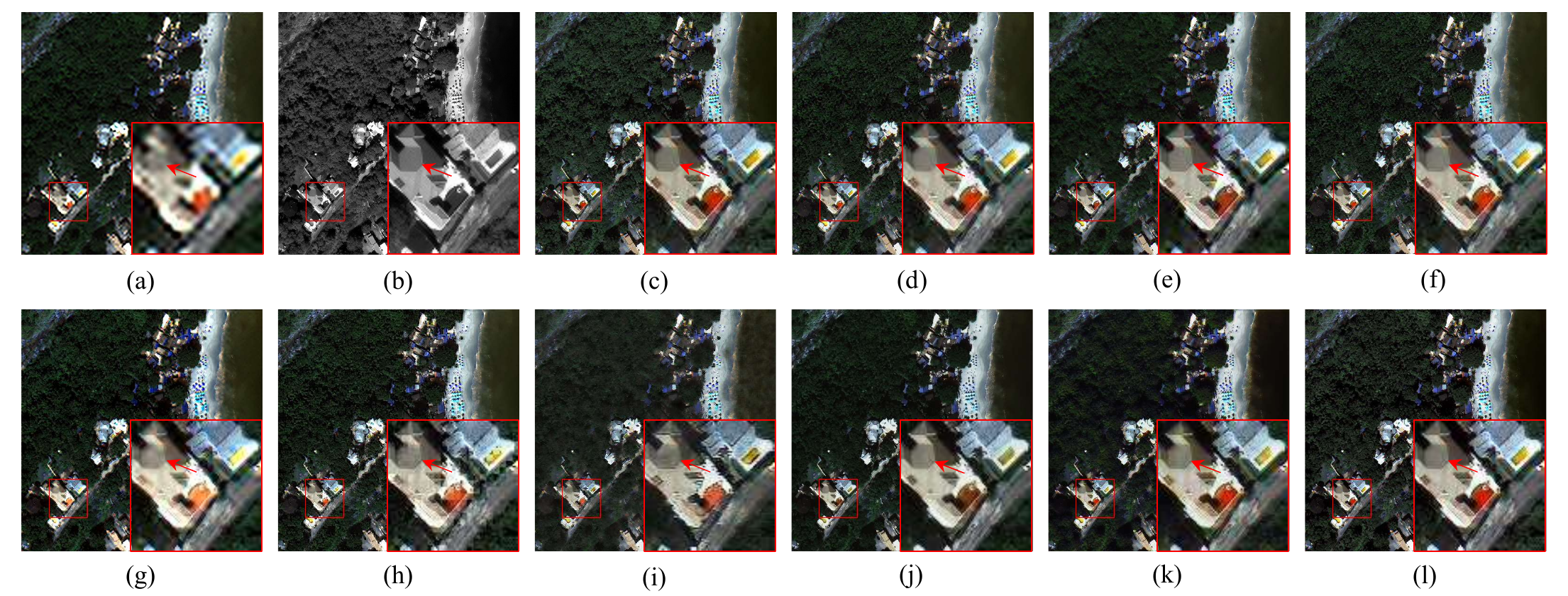}
\end{center}
\vspace{-0.4cm}
\caption{Pansharpened results from different methods on the WV3 dataset at full resolution. (a) Upsampled LRMS. (b) PAN. (c) PCA. (d) HIS. (e) Brovey. (f) GS. (g) BDSD. (h) AWLP. (i) PanNet. (j) PNN. (k) Pan-GAN. (l) Ours.}
\label{figure9}
\end{figure*}
\vspace{-0.4cm}
\begin{figure*}[t]
\begin{center}
\includegraphics[width=6in]{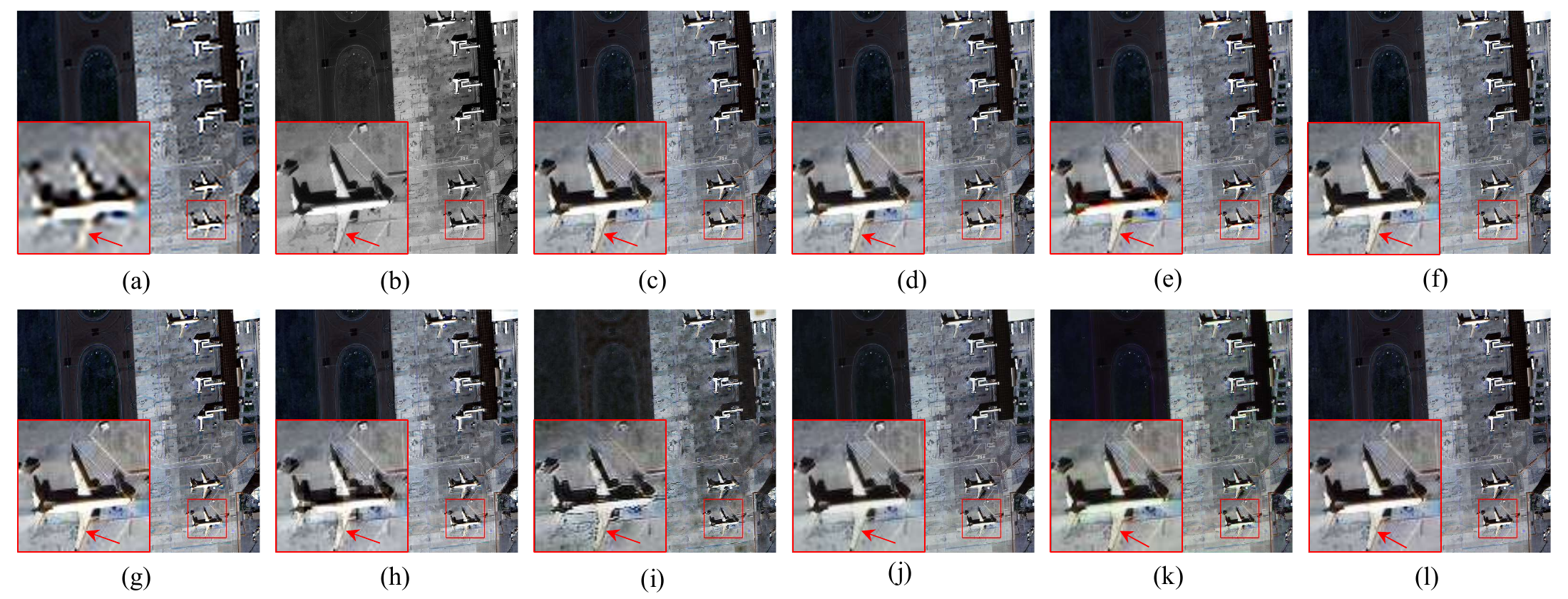}
\end{center}
\vspace{-0.4cm}
\caption{Pansharpened results from different methods on the WV3 dataset at full resolution. (a) Upsampled LRMS. (b)PAN. (c) PCA. (d) HIS. (e) Brovey. (f) GS. (g) BDSD. (h) AWLP. (i) PanNet. (j) PNN. (k) Pan-GAN. (l) Ours.}
\label{figure10}
\end{figure*}
\vspace{-0.4cm}
\begin{figure*}[t]
\begin{center}
\includegraphics[width=6in]{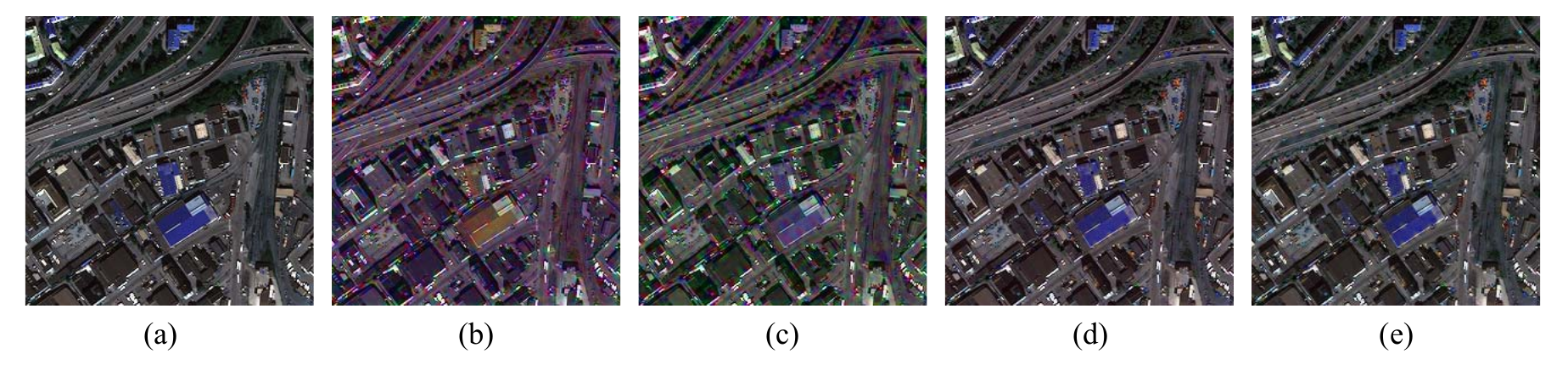}
\end{center}
\vspace{-0.4cm}
\caption{Pansharpened results from the ablation study of the loss functions. (a) Ground truth. (b) Without ${L_{spatial\_l}}$ and ${L_{KL}}$. (c) Without ${L_{KL}}$. (d) Without ${L_{spatial\_l}}$. (e) All losses included.}
\label{figure11}
\end{figure*}


\end{document}